\documentclass[twoside,twocolumn,polish,english,prb, showpacs]{revtex4}
\usepackage[T1]{fontenc}
\usepackage[latin2,latin9]{inputenc}
\usepackage{array}
\usepackage{longtable}
\usepackage{amsbsy}
\usepackage{graphicx}
\usepackage{amssymb}

\makeatletter

\providecommand{\tabularnewline}{\\}

\@ifundefined{textcolor}{}
{%
 \definecolor{BLACK}{gray}{0}
 \definecolor{WHITE}{gray}{1}
 \definecolor{RED}{rgb}{1,0,0}
 \definecolor{GREEN}{rgb}{0,1,0}
 \definecolor{BLUE}{rgb}{0,0,1}
 \definecolor{CYAN}{cmyk}{1,0,0,0}
 \definecolor{MAGENTA}{cmyk}{0,1,0,0}
 \definecolor{YELLOW}{cmyk}{0,0,1,0}
 }

\makeatother

\usepackage{babel}

\begin{document}

\title{Frustration effects in rapidly rotating square and triangular optical
lattices}

\author{T. P. Polak}

\address{Adam Mickiewicz University of Pozna\'{n}, Faculty of Physics, Umultowska
85, 61-614 Pozna\'{n}, Poland}

\author{T. K. Kope\'c}

\address{Institute for Low Temperatures and Structure Research, Polish Academy
of Sciences, POB 1410, 50-950 Wroclaw 2, Poland}
\begin{abstract}
We discuss the ground state of the two-dimensional Bose-Hubbard (BH)
Hamiltonian, relevant for rotating gaseous Bose-Einstein condensates,
by employing $\mathrm{U}\left(1\right)$ quantum rotor approach and
the topologically constrained path integral that includes a summation
over $\mathrm{U}\left(1\right)$ topological charge. We derive an
effective quantum action for the BH model, which enables a non-perturbative
treatment of the zero-temperature phase transition. We calculate the
ground-state phase diagram, analytically deriving maximum repulsive
energy for several rational values of the frustration rotation parameter
$f=0$, $1/2$, $1/3$, $1/4$, and $1/6$ for the square and triangular
lattice, which improves upon previous theoretical treatments. The
ground state of the rotating Bose-Einstein condensates on a triangular
lattice appears to be most stable against the effects of rotation.
Performed calculations revealed strong dependence of the critical
ratio of the kinetic energy to the repulsive on-site energy, that
separates the global coherent from the insulating state, on topology
of the lattice.
\end{abstract}


\maketitle

\section{Introduction}

The merging of atomic and condensed matter physics since the experimental
realization of Bose-Einstein condensation\cite{anderson} has opened
exciting new perspectives for the creation of novel quantum states.
Especially, systems of ultra-cold atoms confined in optical lattices
\cite{jaksch,greiner,greiner1} facilitate an experimental environment,
where a rich variety of quantum many-body models can be implemented
in a wide range of spatial dimensions, geometries, and particle interactions.
Surprisingly, the quantum phase transitions in systems under uniform
magnetic field can be also analyzed considering rotating Bose-Einstein
condensates\cite{coddington,tung,schweikhard} trapped in a two-dimensional
($2D$) lattice potential. In a frame of reference rotating about
the $z$-axis with angular velocity $\Omega$ the kinetic term in
Hamiltonian is equivalent to that of a particle of charge $q$ experiencing
a magnetic field $B$ with $qB=2m\Omega$, where $m$ is the mass
of the particle.\cite{bhat1,bhat2} This connection shows that the
Coriolis force in the rotating frame plays the same role as the Lorentz
force on a charged particle in an uniform magnetic field.\cite{cooper,leggett}
The presence of angular velocity induces vortices in the system described
by the rotation frustration parameter $f$ ($\equiv ma^{2}\Omega/\pi\hbar$,
with $a$ being the lattice spacing). The parameter $f$ can be also
expressed in terms of the recoil energy $E_{R}$ as $f=\pi\hbar\Omega/2E_{R}$.
Of special interest are cases when $f=p/q$, with $p$ and $q$ being
the rational numbers. Frustration occurs in this system because two
different area scales are in competition. One characteristic area
is the unit cell $a$ of considered lattice. The other $\pi\hbar/m\Omega$
is associated with the rotation of the lattice. We can use the notion
\textquotedblleft{}magnetic field\textquotedblright{} and \textquotedblleft{}rotation\textquotedblright{}
interchangeably, assuming that a harmonic confinement potential is
applied to cancel the centrifugal effects of rotation. Therefore,
the nexus of condensed matter and optical physics is transparent (since
the effects of magnetic field/rotation have the same mathematical
structure) and different systems can mimic each other. 

The progress in setups contrivance used for creation of a rotating
optical lattice led to systems with different geometries like square
($\square$) or triangular ($\vartriangle$) that can be analyzed
in the strongly interacting regime.\cite{tung} Up to now two experimental
strategies have been developed.\cite{chevy} The first one is based
on direct imprinting a phase shift on the macroscopic wave function.\cite{matthews}
The second approach called {}``stirring'' is an adaptation of the
rotating bucket experiment to a gas of trapped bosons.\cite{shaer,hodby,haljan}
The latter method fails when angular velocity is comparable to trapping
frequency $\Omega\sim\omega_{\perp}$. However, several groups have
found a way to circumvent the problem of the center of mass expulsion
occurring at $\Omega\sim\omega_{\perp}$ and one can achieve $\Omega=1.05\omega_{\perp}$.\cite{bretin}
Mott-insulator (MI) - superfluid (SF) transition boundary obtained
by using a Gutzwiller-type variational wave function revealed the
complexity of the dependence of phase boundary on the effective magnetic
field/rotation, reflecting the self-similar properties of the single
particle energy spectrum.\cite{umu} Mean field theory calculations
determined that the linear eigenvalue equation characterizing the
Mott lobe also characterizes the Hofstadter butterfly spectrum. From
this authors determined an expression for the Mott-lobe boundary.\cite{goldbaum}
Despite the several theoretical approaches to the problem of strongly
interacting bosons in rotating lattices many questions still remain
open and unsolved.

The aim of this work is to study the superfluid to Mott-insulator
zero-temperature phase transition by means of the Bose-Hubbard (BH)
model in two-dimensional rotating optical condensates with different
geometries. We address the question of evolution of the ground state
phase diagram for the system with various angular velocities for square
and triangular lattices. The inherent difficulty of dealing with BH
Hamiltonian appropriate for strongly correlated bosons originates
from the non-perturbative nature of the model and the presence of
rotation. To elucidate the quantum phase transition in optical lattices,
where the kinetic energy scale is less than the dominating interaction
energy and angular velocity is comparable to the recoil energy, we
have adopted a theoretical approach for strongly interacting fermions
\cite{kopec} to the BH model in a way to include the effects of particle
number fluctuations and make the qualitative phase diagrams more quantitative.\cite{polak}
To facilitate this task, we employ a functional integral formulation
of the theory that enables to perform functional integration over
fields defined on different topologically equivalent classes of the
$\mathrm{U}\left(1\right)$ group, i.e., with different winding numbers.
An inclusion of the winding numbers is unavoidable in order to obtain
a proper phase diagram. The quantum rotor representation method we
use is deeply rooted in the gauge symmetries of the model. We construct
an invariant theory introducing an appropriate $\mathrm{U}\left(1\right)$
gauge transformation. 

The outline of the paper is as follows In Sec. II we introduce the
model Hamiltonian and the effects of rotation are discussed in Sec.
III. Next, we derive an effective $\mathrm{U}\left(1\right)$ action
in the quantum rotor representation described in Sec. IV-VII. The
aim of Sec. VIII is the presentation of the resulting phase diagrams
for two-dimensional square and triangular Bose-Hubbard systems in
rotating frame and comparison of our results with several numerical
and analytical calculations. Finally, Sec. IX summarizes our results.
In the Appendix, we give an analytical derivation of density of states
(DOS) in closed form for several rational values of $f=p/q$ in $\square$
and $\triangle$ lattice. Moreover, the connection between the DOS
and Hofstadter butterfly is shown.

\section{Model}

In optical lattices the two main energy scales are set by the hopping
amplitude proportional to $t$ (that sets the kinetic energy scale
for bosons) due to the particles tunneling, and the on-site interaction
$U>0$. For $t>U$ the phases of the superfluid order parameter on
individual lattice sites are well defined. On the other hand, for
sufficiently large repulsive energy $U$, the quantum phase fluctuations
lead to complete suppression of the long-range phase coherence even
at zero temperature. The competition between the kinetic energy, which
is gained by delocalizing bosons over lattice sites and the repulsive
interaction energy, which disfavors having more than one particle
at any given site, can be modeled by the following quantum Bose-Hubbard
Hamiltonian\cite{fisher}\begin{equation}
\mathcal{H}=\frac{U}{2}\sum_{\boldsymbol{r}}n_{\boldsymbol{r}}\left(n_{\boldsymbol{r}}-1\right)-\sum_{\left\langle \boldsymbol{r,}\boldsymbol{r'}\right\rangle }t_{\boldsymbol{rr'}}a_{\boldsymbol{r}}^{\dagger}a_{\boldsymbol{r'}}-\mu\sum_{\boldsymbol{r}}n_{\boldsymbol{r}},\label{hamiltonian1}\end{equation}
where $a_{\boldsymbol{r}}^{\dagger}$ and $a_{\boldsymbol{r'}}$ stand
for the bosonic creation and annihilation operators that obey the
canonical commutation relations $[a_{\boldsymbol{r}},a_{\boldsymbol{r'}}^{\dagger}]=\delta_{\boldsymbol{r}\boldsymbol{r}'}$,
$n_{\boldsymbol{r}}=a_{\boldsymbol{r}}^{\dagger}a_{\boldsymbol{r}}$
is the boson number operator on the site $\boldsymbol{r}$, and the
chemical potential $\mu$ controls the number of bosons. Here, $\left\langle \boldsymbol{r},\boldsymbol{r'}\right\rangle $
identifies summation over the nearest-neighbor sites. Furthermore,
$t_{\boldsymbol{r}\boldsymbol{r}'}$ is the hopping matrix element
with dispersion $t_{\boldsymbol{k}}^{\square,\triangle}$.

\section{Effects of rotation}

In the fast rotation regime the physics of Bose-Einstein condensates
is very reminiscent of that of charged particle in magnetic field.
If the centrifugal term can be compensated by the trapping frequency
in the plane perpendicular to the rotation axis ($\Omega=\omega_{\perp}$),
so only the Coriolis term is left, we have situation which formally
is equivalent to the Lorentz force exerted by uniform magnetic field
on charged particle. Experimentally a region of fast rotations up
to $\Omega=1.05\omega_{\perp}$\cite{bretin} that suits our theoretical
predictions can be achieved.

An angular velocity enters the Hamiltonian Eq. (\ref{hamiltonian1})
through the Peierls phase factor according to \begin{equation}
t_{\boldsymbol{rr'}}\rightarrow t_{\boldsymbol{rr'}}\exp\left(\frac{2\pi i}{\kappa}\int_{\boldsymbol{r}}^{\boldsymbol{r}'}\boldsymbol{\mathrm{A}}\cdot d\boldsymbol{l}\right),\label{peierls phase factor}\end{equation}
where $\boldsymbol{A}\left(\boldsymbol{r}\right)=\boldsymbol{\Omega}\times\boldsymbol{r}$
is the equivalent of a magnetic vector appears from the rotation and
$\kappa=h/m$ is the quantum circulation unit. Thus, the phase shift
on each site is determined by the vector potential $\boldsymbol{\mathrm{A}}\left(\boldsymbol{r}\right)$
and in typical experimental situations can be entirely ascribed to
the external magnetic field/angular velocity. We assume throughout
this paper that the model in Eq. (\ref{hamiltonian1}) is defined
on a lattice with lattice spacing $a=1$. From Eq. (\ref{peierls phase factor}),
it follows that the properties of the system will be periodic with
a period corresponding to \begin{equation}
f=\pi\hbar\Omega/2\mbox{\ensuremath{E_{R}}}\label{frustration parameter}\end{equation}
per plaquette. Of special interest are the values of the angular momentum
which correspond to rational numbers of $f=1/2,1/3,1/4,...$ Since
all properties of the Hamiltonian Eq. (\ref{hamiltonian1}) are invariant
under $ $$f\rightarrow-f$ and also under $f\rightarrow f+1$, it
is sufficient to consider $f$ in the range $0<f<1/2$ that can be
reached experimentally.

\section{Description of the method}

We write the partition function of the system switching from the particle-number
representation to the conjugate phase representation of the bosonic
degrees of freedom using the bosonic path-integral over the complex
fields $a_{\boldsymbol{r}}\left(\tau\right)$ depending on the {}``imaginary
time'' $0\leq\tau\leq\beta\equiv1/k_{\mathrm{B}}T$ with $T$ being
the temperature\begin{eqnarray}
\mathcal{Z} & = & \int\left[\mathcal{D}\bar{a}\mathcal{D}a\right]\exp\left[-\int_{0}^{\beta}d\tau\mathcal{H\left(\tau\right)}\right.\nonumber \\
 &  & \left.-\sum_{\boldsymbol{r}}\int_{0}^{\beta}d\tau\bar{a}_{\boldsymbol{r}}\left(\tau\right)\frac{\partial}{\partial\tau}a_{\boldsymbol{r}}\left(\tau\right)\right].\end{eqnarray}
 We decouple the interaction term in Eq. (\ref{hamiltonian1}) by
a Gaussian integration over the auxiliary scalar potential fields
\begin{equation}
V_{\boldsymbol{r}}\left(\tau\right)=V_{\boldsymbol{r}0}+V_{\boldsymbol{r}}'\left(\tau\right),\end{equation}
with static \begin{equation}
V_{\boldsymbol{r}0}=\beta^{-1}V_{\boldsymbol{r}}\left(\omega_{\nu}=0\right)\end{equation}
and periodic part \begin{equation}
V'_{\boldsymbol{r}}\left(\tau\right)=\beta^{-1}\sum_{\nu=1}^{+\infty}V_{\boldsymbol{r}}\left(\omega_{\nu}\right)\exp\left(i\omega_{\nu}\tau\right)+\mathrm{c.c},\label{periodic part V}\end{equation}
where $\omega_{\nu}=2\pi\nu/\beta$ ($\nu=0,\pm1,\pm2,...$) is the
Bose-Matsubara frequency. We observe now that the BH Hamiltonian has
a local $\mathrm{U}\left(1\right)$ gauge symmetry, when expressed
in terms of the underlying boson variables. This points out a possibility
of an emergent dynamical $\mathrm{U}\left(1\right)$ gauge field as
a fluctuating complex field attached to bosonic variables, which is
dynamically generated, by interacting bosons. Thus, the periodic part
$V'_{\boldsymbol{r}}\left(\tau\right)\equiv V'_{\boldsymbol{r}}\left(\tau+\beta\right)$
couples to the local particle number through the Josephson-like relation
$\dot{\phi}_{\boldsymbol{r}}\left(\tau\right)=V'_{\boldsymbol{r}}\left(\tau\right)$,
where \begin{equation}
\dot{\phi}_{\boldsymbol{r}}\left(\tau\right)\equiv\frac{\partial\phi_{\boldsymbol{r}}\left(\tau\right)}{\partial\tau}=e^{-\phi_{\boldsymbol{r}}\left(\tau\right)}\frac{1}{i}\frac{\partial}{\partial\tau}e^{\phi_{\boldsymbol{r}}\left(\tau\right)}.\end{equation}
The quantity $\phi\left(\tau\right)$ is the $\mathrm{U}\left(1\right)$
\textit{phase} field and satisfies the periodicity condition $\phi_{\boldsymbol{r}}\left(\beta\right)=\phi_{\boldsymbol{r}}\left(0\right)$
as a consequence of the periodic properties of the $V'_{\boldsymbol{r}}\left(\tau\right)$
field in Eq. (\ref{periodic part V}).

\section{Parametrization of the boson field and the order parameter}

We perform the local gauge transformation to the \textit{new} bosonic
variables \begin{equation}
\left[\begin{array}{c}
a_{\boldsymbol{r}}\left(\tau\right)\\
\bar{a}_{\boldsymbol{r}}\left(\tau\right)\end{array}\right]=\left[\begin{array}{cc}
e^{i\phi_{\boldsymbol{r}}\left(\tau\right)} & 0\\
0 & e^{-i\phi_{\boldsymbol{r}}\left(\tau\right)}\end{array}\right]\left[\begin{array}{c}
b_{\boldsymbol{r}}\left(\tau\right)\\
\bar{b}_{\boldsymbol{r}}\left(\tau\right)\end{array}\right],\label{gauge}\end{equation}
that removes the imaginary term $-i\int_{0}^{\beta}d\tau\dot{\phi}_{\boldsymbol{r}}\left(\tau\right)n_{\boldsymbol{r}}\left(\tau\right)$
from all the Fourier modes except at zero frequency. From the above
we deduce bosons have a composite nature made of bosonic part $b_{\boldsymbol{r}}\left(\tau\right)$
and attached {}``flux'' $\exp\left[i\phi_{\boldsymbol{r}}\left(\tau\right)\right]$.
We parametrize the boson fields $b_{\boldsymbol{r}}\left(\tau\right)=b_{0}+b_{\boldsymbol{r}}^{'}\left(\tau\right)$
and incorporate fully our calculations to the phase fluctuations governed
by the gauge group $\mathrm{U}\left(1\right)$. Assuming nonfluctuating
amplitude at low temperatures $b_{\boldsymbol{r}}\left(\tau\right)=b_{0}$,
we drop the corrections, which was proved to be justified in the large
$U/t$ limit we are interested in.\cite{polak,kampf} The calculation
of $b_{0}$ is postponed to the next section.

\begin{figure}
\includegraphics[scale=0.4]{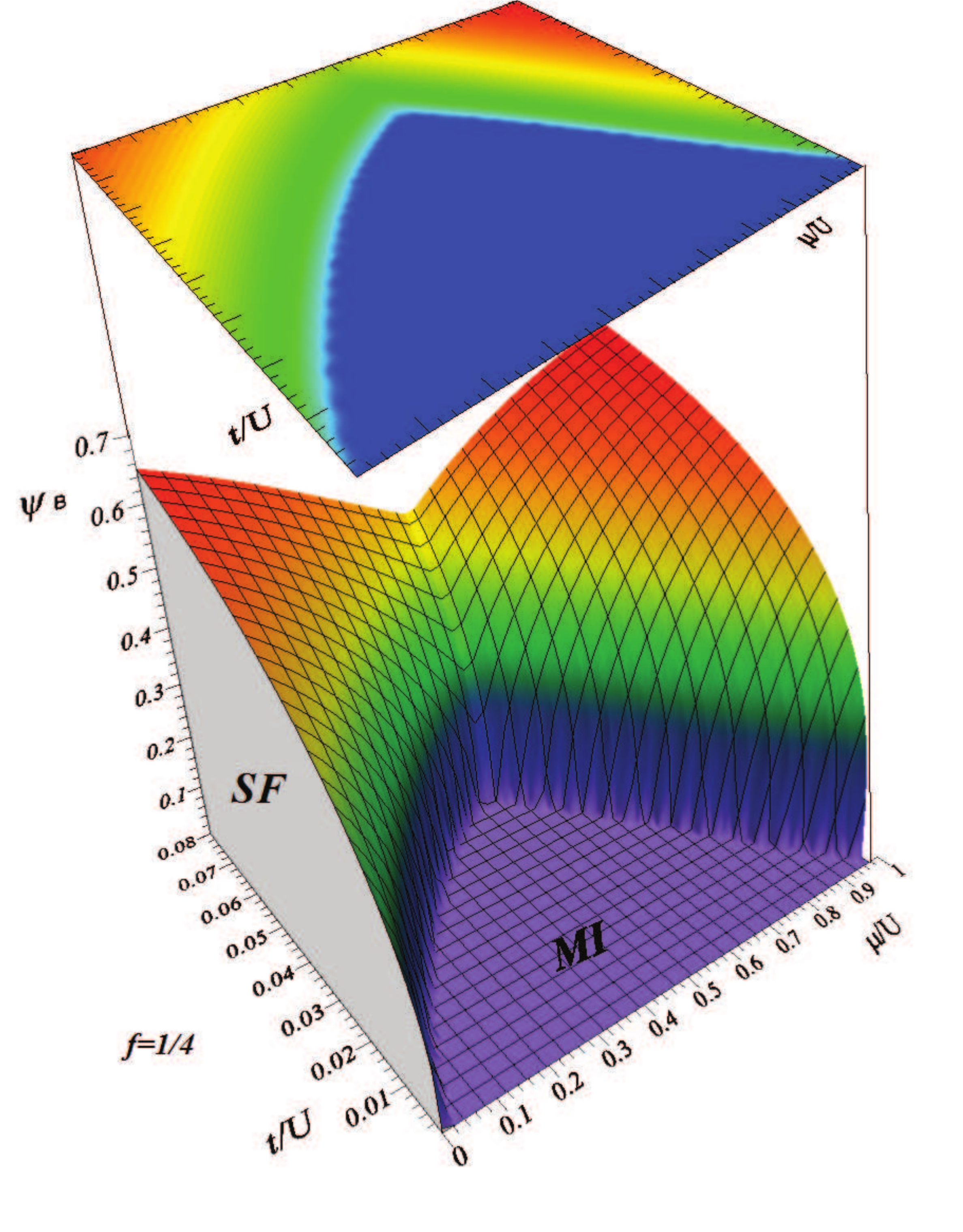}\caption{(Color online) The phase variable's part $\psi_{B}$ of the order
parameter $\Psi_{B}$ Eq. (\ref{order parameter definition}) for
rotating triangular lattice with the rotation frustration parameter
$f=1/4$. Below the surface the phase coherent state (SF) takes place.
Flat region means the incompressible Mott-insulator (MI) phase. The
upper panel is the density plot of the surface in order to highlight
the interaction $t/U$- chemical potential $\mu/U$ dependence.}
\label{order diagram 14}
\end{figure}

\begin{figure}
\includegraphics[scale=0.7]{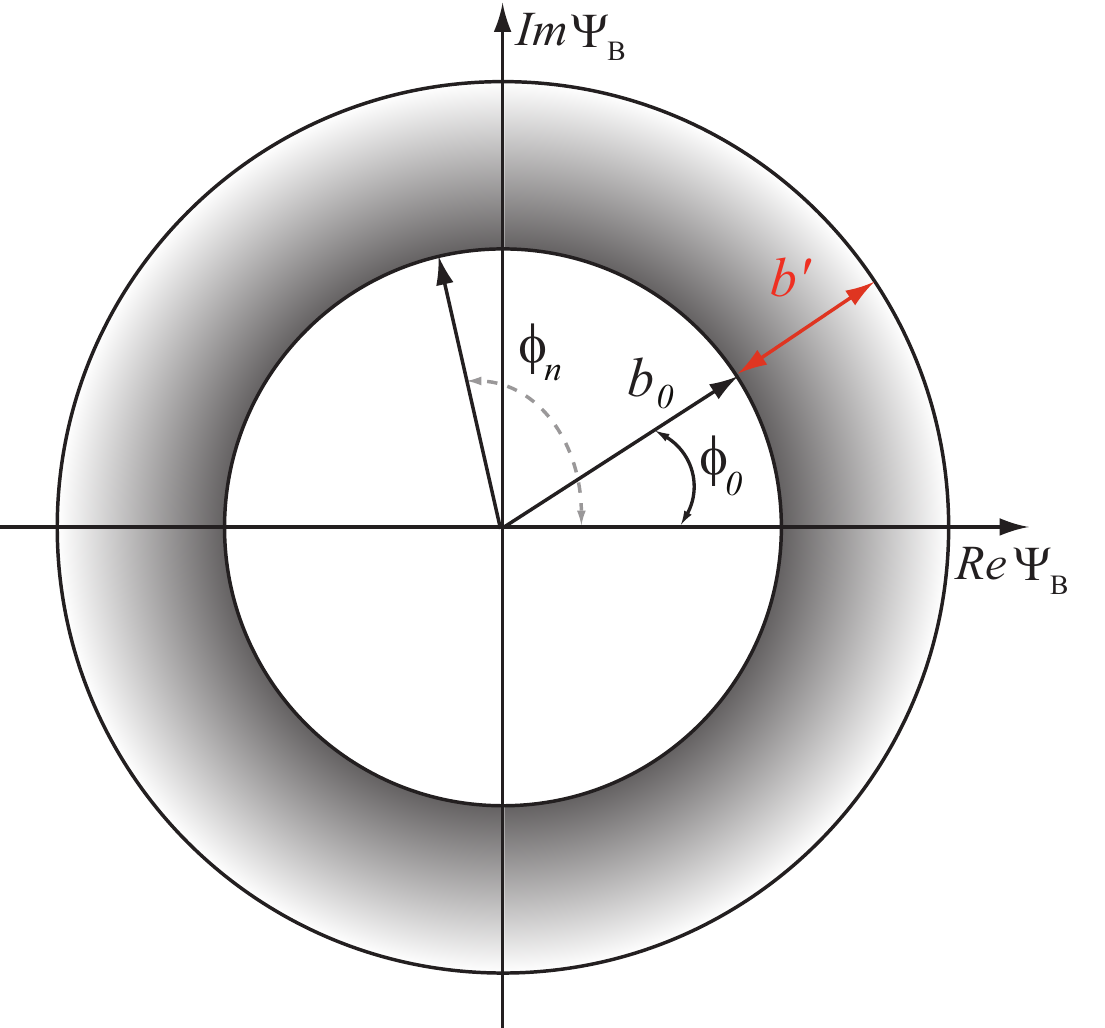}\caption{(Color online) The order parameter $\Psi_{B}$ can achieve nonzero
value when \emph{both} amplitude $b_{0}$ and phase $\phi$ are positive.
The fluctuations of the amplitude $b'$ in the Matsubara time in low-temperature
limit $T\rightarrow0$ are dropped in our approach.}
\label{order concept}
\end{figure}
It is very convenient to define the order parameter \begin{equation}
\Psi_{B}\equiv\left\langle a_{\boldsymbol{r}}\left(\tau\right)\right\rangle =\left\langle b_{\boldsymbol{r}}\left(\tau\right)\exp\left[i\phi_{\boldsymbol{r}}\left(\tau\right)\right]\right\rangle =b_{0}\psi_{B},\label{order parameter definition}\end{equation}
which signals the emergence of the superfluid phase and vanishes in
the Mott-insulator state. The SF state is characterized by spontaneously
breaking of the $\mathrm{U\left(1\right)}$ symmetry of the Bose-Hubbard
Hamiltonian. Note, that a nonzero value of the amplitude $b_{0}$
in Eq. (\ref{order parameter definition}) is \emph{not sufficient}
for superfluidity. To achieve this, also the phase variables $\phi$
in Eq. (\ref{order parameter definition}), must become stiff and
coherent, which implies $\psi_{B}\neq0$ (see Fig. \ref{order diagram 14}
and Fig. \ref{order concept}). In the symmetry breaking state, with
a finite expectation value of $a_{\boldsymbol{r}}\left(\tau\right)$,
different phases $\phi_{1},\phi_{2},...,\phi_{n}$ of the condensate
lead to degenerate ground states (Fig. \ref{order concept}). If we
change the phase of the condensate in a large but finite region, then,
locally, the system is still in one of the degenerate ground states.
Slow changes of the phase result in the appearance of the low energy
excitations that correspond to fluctuations among the degenerate states.
Based on this picture we identify $b'$ as high energy fast fluctuations
(and drop them in calculations) contrary to the low energy fluctuations
described by $\phi$.

\section{Phase only action}

By integrating out the auxiliary static field $V_{\boldsymbol{r}0}$
we calculate the partition function with an effective action expressed
in the form of the propagator $\hat{G}$\begin{equation}
\mathcal{Z}=\int\left[\mathcal{D}\phi\right]e^{-\sum_{\boldsymbol{r}}\int_{0}^{\beta}d\tau\left[\frac{1}{2U}\dot{\phi}_{\boldsymbol{r}}^{2}\left(\tau\right)+\frac{1}{i}\frac{\bar{\mu}}{U}\dot{\phi}_{\boldsymbol{r}}\left(\tau\right)\right]+\mathrm{Tr}\ln\hat{G}^{-1}},\label{partition function propagator}\end{equation}
where $\bar{\mu}/U=\mu/U+1/2$ is the shifted reduced chemical potential.
In the above $\exp\left(-\mathrm{Tr}\ln\hat{G}^{-1}\right)\equiv\det\hat{G}$
and the determinant takes the form\begin{eqnarray}
\det\hat{G} & = & \int\left[\mathcal{D}\bar{b}\mathcal{D}b\right]\exp\left\{ -\sum_{\left\langle \boldsymbol{r},\boldsymbol{r'}\right\rangle }\int_{0}^{\beta}d\tau\right.\nonumber \\
 &  & \times\bar{b}_{\boldsymbol{r}}\left[\left(\frac{\partial}{\partial\tau}+\bar{\mu}\right)\delta_{\boldsymbol{r}\boldsymbol{r}'}\right.\nonumber \\
 &  & -\left.\left.e^{i\phi_{\boldsymbol{r}}\left(\tau\right)}t_{\boldsymbol{r}\boldsymbol{r'}}e^{-i\phi_{\boldsymbol{r'}}\left(\tau\right)}\right]b_{\boldsymbol{r}}\right\} .\end{eqnarray}
The inverse of the propagator becomes\begin{equation}
\hat{G}^{-1}=\hat{G}_{0}^{-1}-K=\hat{G}_{0}^{-1}\left(1-K\hat{G}_{0}\right).\end{equation}
The explicit value of the amplitude $b_{0}$ in Eq. (\ref{order parameter definition})
can be obtained from minimization of the Hamiltonian $\partial\mathcal{H}\left(b_{0}\right)/\partial b_{0}=0$.
Therefore, we write\begin{eqnarray}
\hat{G}_{0} & = & b_{0}^{2}\equiv\frac{\sum_{\left\langle \boldsymbol{r},\boldsymbol{r'}\right\rangle }t_{\boldsymbol{r}\boldsymbol{r'}}+\bar{\mu}}{U},\label{zero mode}\\
K & = & e^{i\phi_{\boldsymbol{r}}\left(\tau\right)}t_{\boldsymbol{r}\boldsymbol{r'}}e^{-i\phi_{\boldsymbol{r'}}\left(\tau\right)}.\label{hopping}\end{eqnarray}
Expanding the trace of the logarithm we have \begin{eqnarray}
\mathrm{Tr}\ln\hat{G}^{-1} & = & -\mathrm{Tr}\left(\ln\hat{G}_{0}\right)-\mathrm{Tr}\left(K\hat{G}_{0}\right)\nonumber \\
 &  & -\frac{1}{2}\mathrm{Tr}\left[\left(K\hat{G}\right)^{2}\right]+...\end{eqnarray}
with $\hat{G}_{0}$ and $K$ given by Eq. (\ref{zero mode}) and (\ref{hopping}).
Finally the partition function Eq. (\ref{partition function propagator})
becomes\begin{eqnarray}
\mathcal{Z} & = & \int\left[\mathcal{D}\phi\right]e^{-\mathcal{S}_{\mathrm{phase}}\left[\phi\right]}\end{eqnarray}
with an effective action expressed \emph{only} in the \emph{phase}
fields variable \begin{eqnarray}
\mathcal{S}_{\mathrm{phase}}\left[\phi\right] & = & \int_{0}^{\beta}d\tau\left\{ \sum_{\boldsymbol{r}}\left[\frac{1}{2U}\dot{\phi}_{\boldsymbol{r}}^{2}\left(\tau\right)+\frac{1}{i}\frac{\bar{\mu}}{U}\dot{\phi}_{\boldsymbol{r}}\left(\tau\right)\right]\right.\nonumber \\
 &  & \left.-\sum_{\left\langle \boldsymbol{r},\boldsymbol{r'}\right\rangle }e^{i\phi_{\boldsymbol{r}}\left(\tau\right)}J_{\boldsymbol{r}\boldsymbol{r'}}e^{-i\phi_{\boldsymbol{r'}}\left(\tau\right)}\right\} ,\label{action only phase}\end{eqnarray}
where the phase stiffness coefficient is given by $J_{\boldsymbol{r}\boldsymbol{r'}}=b_{0}^{2}t_{\boldsymbol{r}\boldsymbol{r'}}$.
The total time derivative Berry phase imaginary term in Eq. (\ref{action only phase})
is nonzero due to topological phase field configurations with $\phi_{\boldsymbol{r}}\left(\beta\right)-\phi_{\boldsymbol{r}}\left(0\right)=2\pi m_{\boldsymbol{r}}$
($m_{\boldsymbol{r}}=0,\pm1,\pm2...$) that result in topological
ingredients to the correlator we will see below. Therefore, we concentrate
on closed paths in the imaginary time $\left(0,\beta\right)$ labeled
by the integer winding numbers $m_{\boldsymbol{r}}$. The path-integral
\begin{equation}
\int\left[\mathcal{D}\phi\right]...\equiv\sum_{\left\{ m_{\boldsymbol{r}}\right\} }\int_{0}^{2\pi}\left[\mathcal{D}\phi\left(0\right)\right]\int_{_{\phi_{\boldsymbol{r}}\left(0\right)}}^{\phi_{\boldsymbol{r}}\left(\tau\right)+2\pi m_{\boldsymbol{r}}}\left[\mathcal{D}\phi\left(\tau\right)\right]...,\end{equation}
includes a summation over $m_{\boldsymbol{r}}$ and in each topological
sector the integration goes over the gauge potentials.

To proceed, we replace the phase degrees of freedom by the unimodular
scalar complex field $\psi_{\boldsymbol{r}}$ which satisfies the
quantum periodic boundary condition $\psi_{\boldsymbol{r}}\left(\beta\right)=\psi_{\boldsymbol{r}}\left(0\right)$.
This can be conveniently done using the Fadeev-Popov method with Dirac
delta functional resolution of unity \cite{kopec1}, where we take
$\psi_{\boldsymbol{r}}$ as a continuous but constrained (on the average)
variable to have the unimodular value\begin{eqnarray}
1 & = & \int\left[\mathcal{D}\psi\mathcal{D}\bar{\psi}\right]\delta\left(\sum_{i}\left|\psi_{\boldsymbol{r}}\left(\tau\right)\right|^{2}-N\right)\nonumber \\
 & \times & \prod_{\boldsymbol{r}}\delta\left(\psi_{\boldsymbol{r}}-e^{i\phi_{\boldsymbol{r}}\left(\tau\right)}\right)\delta\left(\bar{\psi}_{\boldsymbol{r}}-e^{-i\phi_{\boldsymbol{r}}\left(\tau\right)}\right),\label{popov}\end{eqnarray}
where $N$ is the number of lattice sites. Introducing the Lagrange
multiplier $\lambda$, which adds the quadratic terms (in the $\psi_{\boldsymbol{r}}$
fields) to the action Eq. (\ref{action only phase}), we can solve
for the constraint. The partition function can be rewritten to the
form\begin{eqnarray}
\mathcal{Z} & = & \int_{-i\infty}^{+i\infty}\left[\frac{\mathcal{D}\lambda}{2\pi i}\right]e^{-N\beta\mathcal{F}\left(\lambda\right)},\end{eqnarray}
 with the free energy density $\mathcal{F}=-\ln\mathcal{Z}/\beta N$
given by:\begin{eqnarray}
\mathcal{F} & = & -\lambda-\frac{1}{N\beta}\ln\int\left[\mathcal{D}\psi\mathcal{D}\bar{\psi}\right]\exp\left\{ \sum_{\boldsymbol{rr'}}\int_{0}^{\beta}d\tau d\tau^{'}\right.\nonumber \\
 &  & \times\left[\left(J\mathcal{I}_{\boldsymbol{rr'}}+\lambda\delta_{\boldsymbol{rr'}}\right)\delta\left(\tau-\tau'\right)\right.\nonumber \\
 &  & -\left.\left.\mathcal{\gamma}_{\boldsymbol{rr'}}\left(\tau,\tau'\right)\right]\psi_{\boldsymbol{r}}\bar{\psi}_{\boldsymbol{r'}}\right\} ,\label{free energy}\end{eqnarray}
where $\mathcal{I}_{\boldsymbol{rr'}}=1$ if $\boldsymbol{r},\boldsymbol{r'}$
are the nearest neighbors and equals zero otherwise, and \begin{equation}
\gamma_{\boldsymbol{rr'}}\left(\tau,\tau'\right)=\left\langle \exp\left\{ -i\left[\phi_{\boldsymbol{r}}\left(\tau\right)-\phi_{\boldsymbol{r'}}\left(\tau'\right)\right]\right\} \right\rangle \end{equation}
 is the two-point phase correlator associated with the order parameter
field, where $\left\langle ...\right\rangle $ denotes averaging with
respect to the action in Eq. (\ref{action only phase}). The final
form of the correlator with topological contribution (summation over
integer winding numbers)\begin{eqnarray}
\gamma_{\boldsymbol{r}\boldsymbol{r}'}\left(\tau,\tau'\right) & = & \frac{\delta_{\boldsymbol{r}\boldsymbol{r}'}\exp\left(\frac{U}{2}\left|\tau-\tau^{'}\right|\right)}{\sum_{\left\{ m_{\boldsymbol{r}}\right\} }\exp\left[-\frac{U\beta}{2}\left(m_{\boldsymbol{r}}+\frac{\bar{\mu}}{U}\right)^{2}\right]}\nonumber \\
 & \times & \sum_{\left\{ m_{\boldsymbol{r}}\right\} }\left\{ \exp\left[-\frac{U\beta}{2}\left(m_{\boldsymbol{r}}+\frac{\bar{\mu}}{U}\right)^{2}\right]\right.\nonumber \\
 & \times & \left.\exp\left[-U\left(m_{\boldsymbol{r}}+\frac{\bar{\mu}}{U}\right)\left(\tau-\tau'\right)\right]\right\} \label{topological contribution}\end{eqnarray}
after Fourier transform, can be written as\begin{equation}
\gamma\left(\omega_{\nu}\right)=\frac{1}{\mathcal{Z}_{0}}\frac{4}{U}\sum_{m=-\infty}^{+\infty}\frac{\exp\left[-\frac{U\beta}{2}\left(m+\frac{\bar{\mu}}{U}\right)^{2}\right]}{1-4\left(m+\frac{\bar{\mu}}{U}-i\frac{\omega_{\nu}}{U}\right)^{2}},\label{correlator}\end{equation}
where \begin{equation}
\mathcal{Z}_{0}=\sum_{m=-\infty}^{+\infty}\exp\left[-U\beta\left(m+\bar{\mu}/U\right)^{2}/2\right]\end{equation}
is the partition function for the set of quantum rotors. The form
of Eq. (\ref{correlator}) assures the periodicity in the imaginary
time with respect to $\bar{\mu}/U=\mu/U+1/2$ which emphasizes the
special role of its integer values. The action Eq. (\ref{action only phase}),
with the topological contribution Eq. (\ref{topological contribution}),
after Fourier transform, is written as \begin{equation}
\mathcal{S}_{\mathrm{eff}}\left[\psi,\bar{\psi}\right]=\frac{1}{N\beta}\sum_{\mathbf{k}\nu}\bar{\psi}_{\mathbf{k}\nu}\mathrm{\Gamma}_{\mathbf{k}}^{-1}\left(\omega_{\nu}\right)\psi_{\mathbf{k}\nu},\end{equation}
where $\mathrm{\Gamma}_{\mathbf{k}}^{-1}\left(\omega_{\nu}\right)=\lambda-J_{\mathbf{k}}+\gamma^{-1}\left(\omega_{\nu}\right)$
is the inverse of the propagator.

\section{Critical Lines}

Within the phase coherent state the order parameter $\psi_{B}$ is
evaluated in the thermodynamic limit $N\rightarrow\infty$ by the
saddle point method $\delta\mathcal{F}/\delta\lambda=0$ and the unimodular
condition of the $\mathrm{U}\left(1\right)$ phase variables translates
into the equation\begin{equation}
1-\psi_{B}^{2}=\frac{1}{N\beta}\sum_{\mathbf{k},\nu}\frac{1}{\lambda-J_{\mathbf{k}}+\gamma^{-1}\left(\omega_{\nu}\right)}.\label{critical line}\end{equation}
The phase boundary is determined by the divergence of the order parameter
susceptibility $\Gamma_{\mathbf{k}=0}\left(\omega_{\nu=0}\right)=0$\begin{equation}
\lambda_{0}-J_{\mathrm{p/q}}^{max}+\gamma^{-1}\left(\omega_{\nu=0}\right)=0\label{lagrange 0}\end{equation}
which determines the critical value of the Lagrange parameter $\lambda=\lambda_{0}$,
that stays constant in the whole global coherent phase. To proceed,
it is desirable to introduce the density of states for a $2D$ lattice
in the rotating frame in the form \begin{equation}
\rho_{p/q}^{\square,\vartriangle}\left(\xi\right)=\frac{1}{N}\sum_{\boldsymbol{k}}\delta\left(\xi-\frac{t_{\boldsymbol{k}}^{\square,\triangle}}{t}\right)\label{DOS definition}\end{equation}
with $t_{\boldsymbol{k}}^{\square,\triangle}$ being the Fourier transform
of the hopping matrix elements. In this context the quantity $J_{\mathrm{p/q}}^{max}$
in Eq. (\ref{lagrange 0}) represents the maximum of the spectrum
described by the DOS Eq. (\ref{DOS definition}). The problem of computing
of $\rho_{p/q}^{\square,\vartriangle}\left(\xi\right)$ reduces effectively
to the solution of the Harper equation relevant, e.g., to tight binding
electrons on a two-dimensional lattice with an uniform magnetic flux
per unit plaquette. In the Appendix, we give an analytical derivation
of $\rho_{p/q}^{\square,\vartriangle}\left(\xi\right)$ in closed
form for several rational values of $p/q$. With the help of the above
and after summation over Bose-Matsubara frequency $\omega_{\nu}$,
the superfluid state order parameter becomes\begin{equation}
1-\psi_{B}^{2}=\frac{1}{2}\int_{-\infty}^{+\infty}\frac{\rho_{p/q}^{\square,\vartriangle}\left(\xi\right)d\xi}{\sqrt{2\bar{\xi}\left(2z\frac{t}{U}+\frac{\mu}{U}+\frac{1}{2}\right)\frac{t}{U}+\upsilon^{2}\left(\frac{\mu}{U}\right)}}.\label{critical line final}\end{equation}
In Eq. (\ref{critical line final}) $\upsilon\left(\mu/U\right)=\mathrm{frac}\left(\mu/U\right)-1/2,$
where $\mathrm{frac}\left(x\right)=x-\left[x\right]$ is the fractional
part of the number and $\left[x\right]$ is the floor function which
gives the greatest integer less than or equal to $x$; $\bar{\xi}=J_{\mathrm{p/q}}^{max}-\xi$
with $J_{\mathrm{p/q}}^{max}$ stands for the maximum value of the
dispersion spectrum $t_{\boldsymbol{k}}^{\square,\triangle}$ and
$z$ is the lattice coordination number.

\section{Phase diagrams}

\begin{figure}
\includegraphics[scale=0.8]{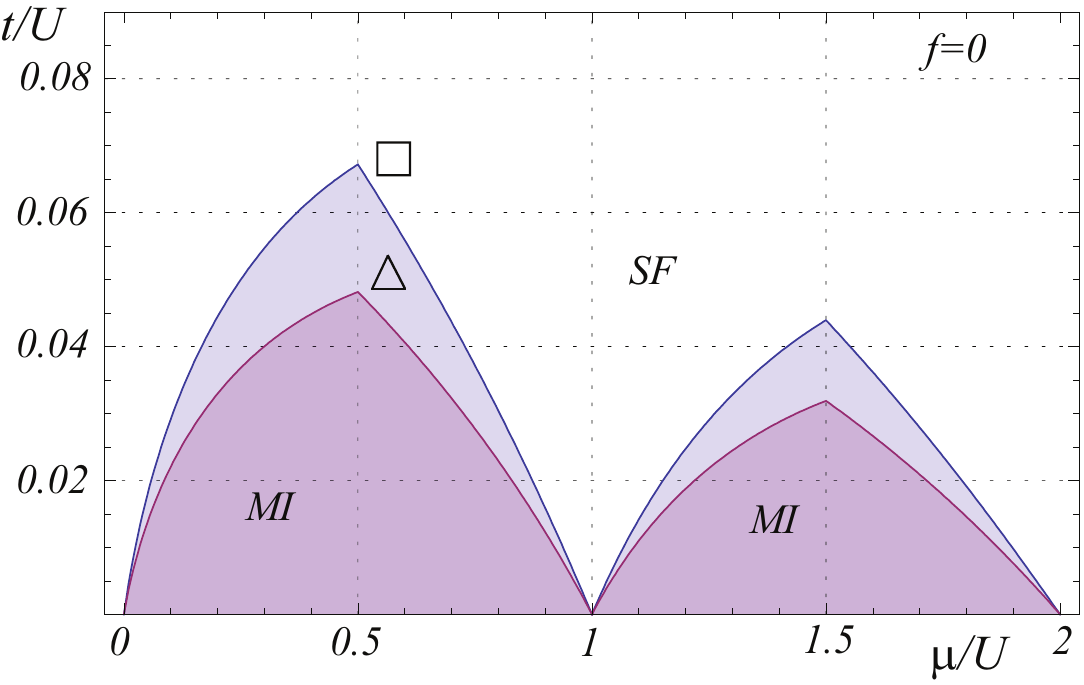}\caption{(Color online) Phase diagram for square $\square$ and triangular
$\bigtriangleup$ lattice (number of particles per lattice site is
$n_{B}=1$ inside the first and $n_{B}=2$ inside the second lobe
respectively) with no rotation $f=0$. Within the lobes the MI phase
takes place with $\Psi_{B}=0$ (see also Fig. \ref{order diagram 14}).}
\label{phase diagram 2dts}
\end{figure}

\begin{figure}
\includegraphics[scale=0.8]{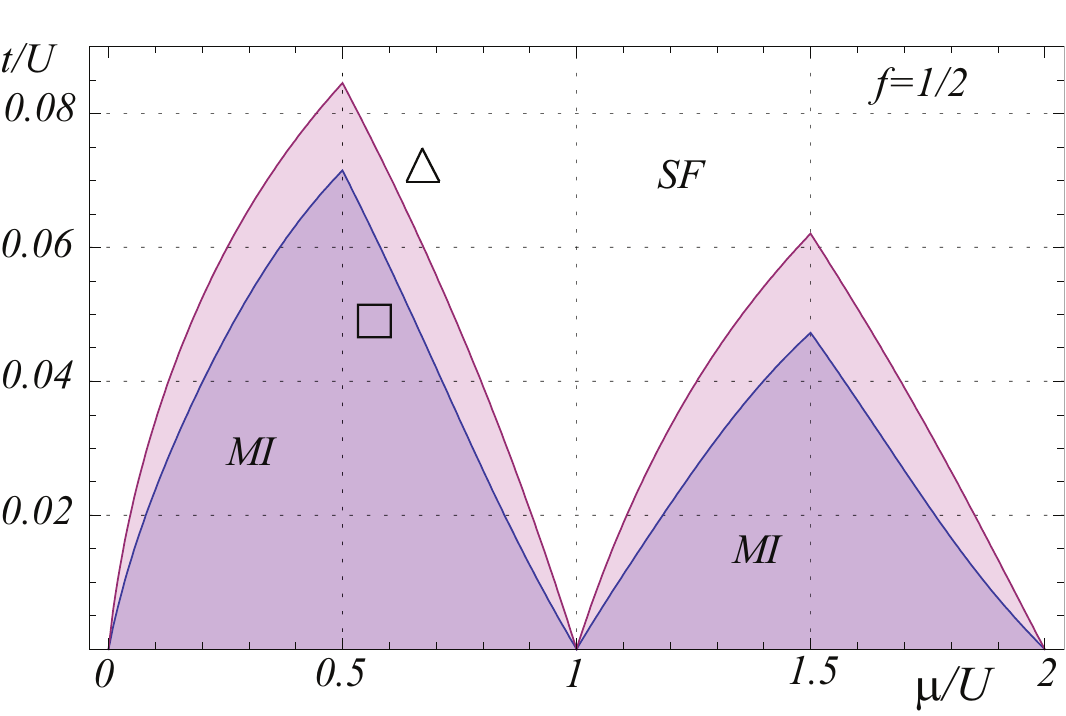}\caption{(Color online) Phase diagram for square $\square$ and triangular
$\bigtriangleup$ lattice (number of particles per lattice site is
$n_{B}=1$ inside the first and $n_{B}=2$ inside the second lobe
respectively) with rotation frustration parameter $f=1/2$. Within
the lobes the MI phase takes place with $\Psi_{B}=0$ (see also Fig.
\ref{order diagram 14}).}
\label{phase diagram 2dts 12}
\end{figure}

\begin{figure}
\includegraphics[scale=0.8]{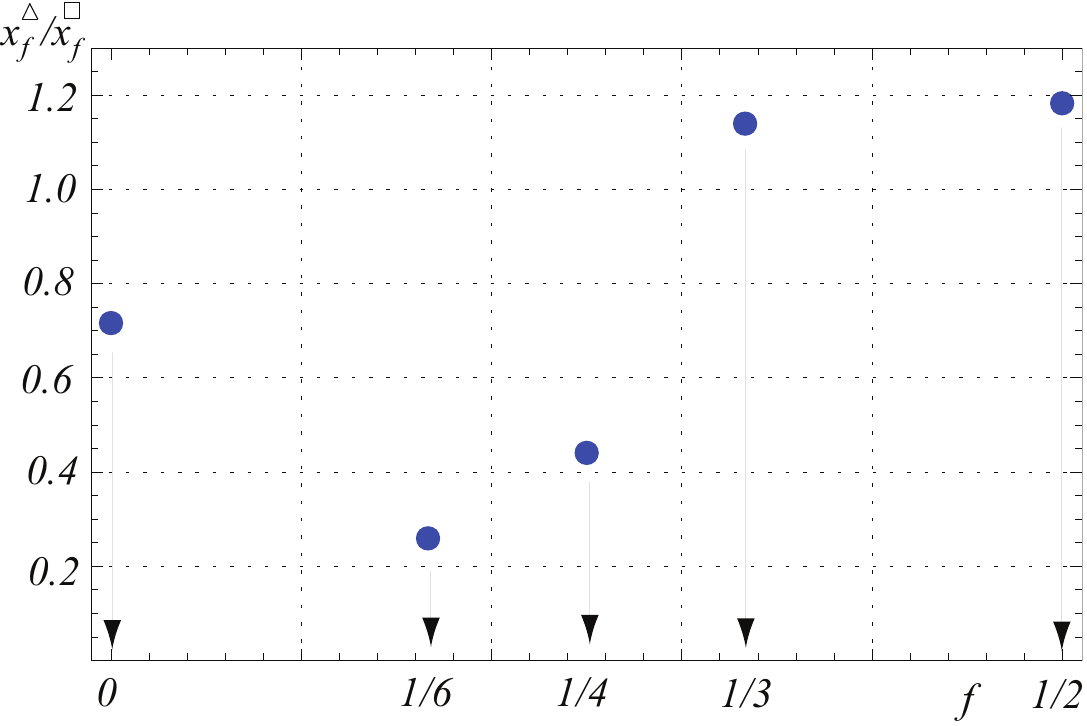}\caption{(Color online) The maximum of the critical value for $t/U$ parameter
(as a function of the normalized chemical potential $\mu/U$) at the
tip of the first ($n_{B}=1$) MI lobe $x_{f}^{\bigtriangleup}/x_{f}^{\square}$
for rotating triangular to square lattice.}
\label{triangular square phase}
\end{figure}

\begin{figure}
\includegraphics[scale=0.8]{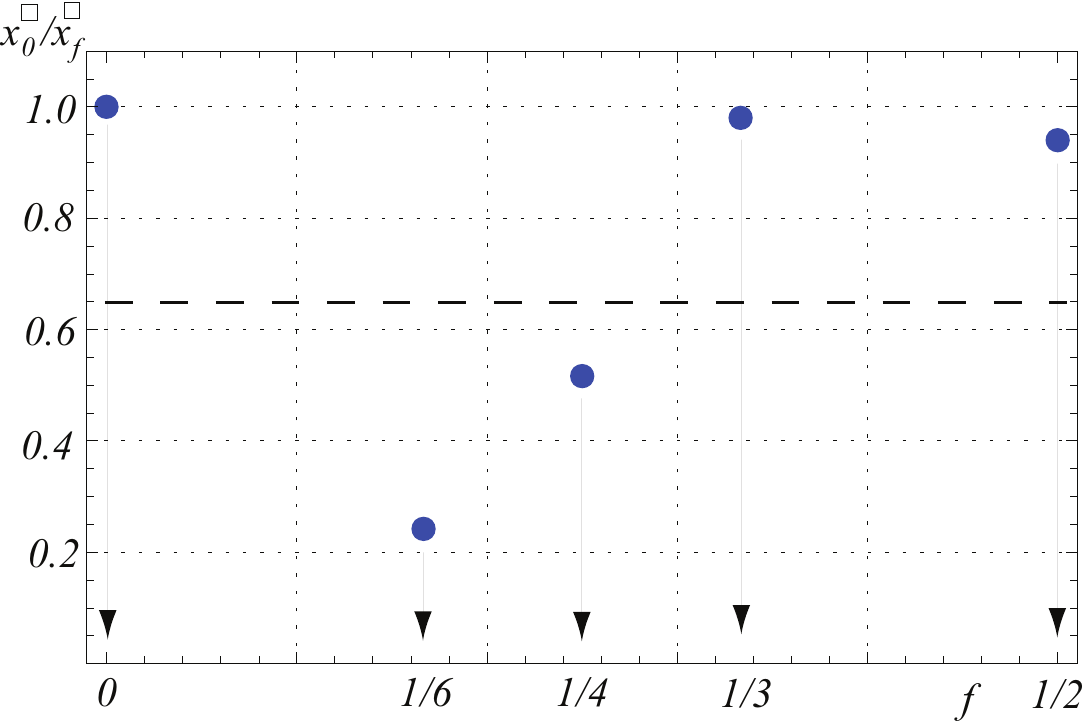}\caption{(Color online) The maximum of the critical value for $t/U$ parameter
(as a function of the normalized chemical potential $\mu/U$) at the
tip of the first ($n_{B}=1$) MI lobe $x_{0}^{\square}/x_{f}^{\square}$
for rotating square lattice. The vertical dashed line marks the ratio
of the maximum of the critical value for $t/U$ parameter for the
second to first lobe $x_{0}^{\square}\left(n_{B}=2\right)/x_{0}^{\square}\left(n_{B}=1\right)$.}
\label{square phase}
\end{figure}

\begin{figure}
\includegraphics[scale=0.8]{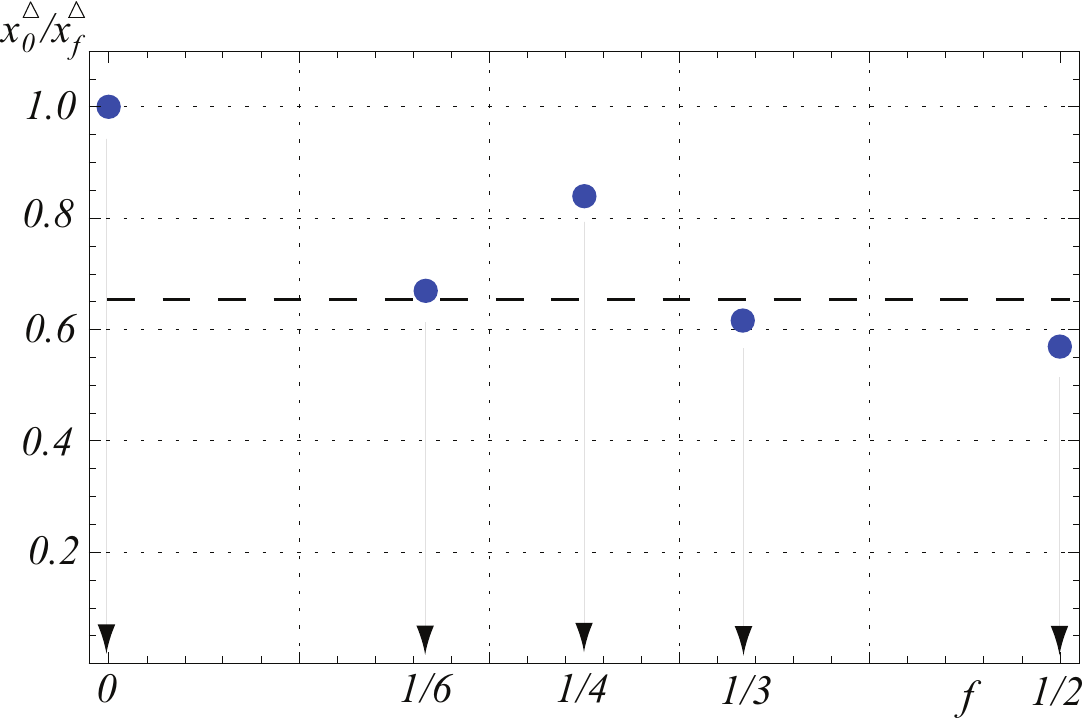}\caption{(Color online) The maximum of the critical value for $t/U$ parameter
(as a function of the normalized chemical potential $\mu/U$) at the
tip of the first ($n_{B}=1$) MI lobe $x_{0}^{\triangle}/x_{f}^{\triangle}$
for rotating triangular lattice. The vertical dashed line marks the
ratio of the maximum of the critical value for $t/U$ parameter for
the second to first lobe $x_{0}^{\triangle}\left(n_{B}=2\right)/x_{0}^{\triangle}\left(n_{B}=1\right)$.}
\label{triangular phase-1}
\end{figure}

$ $The zero-temperature phase diagram of the homogeneous Bose-Hubbard
model Eq. (\ref{hamiltonian1}) can be calculated from Eq. (\ref{critical line final})
and is shown schematically in Fig. \ref{phase diagram 2dts} as a
function of $t/U$, with the density controlled by a chemical potential
$\mu/U$. At $U/t\rightarrow$0, the kinetic energy dominates and
the ground state is a delocalized superfluid, described by nonzero
value of the superfluid order parameter $\Psi_{B}\neq0$. At small
values of $t/U$, interactions dominate and one obtains a series of
MI lobes with fixed integer filling $n_{B}=1,2,...$\cite{polak,fisher}
\begin{table}

\begin{longtable}{c|c|c|c|c|c}
\hline
\hline 
 & QMC & DPT & MFT & PA & QRA\tabularnewline
\endhead
\hline 
$x_{0}^{\square}$ & $0.05974(3)$ & $0.05909$ & $0.043$ & $0.059$ & $0.06719$\tabularnewline
\hline
\hline
\endlastfoot
\end{longtable}\caption{Comparison of the maximum of the critical value for $t/U$ (as a function
of the normalized chemical potential $\mu/U$) at the tip of the first
($n_{B}=1$) MI lobe for the square lattice with several numerical
(QMC - quantum Monte-Carlo,\cite{capogrosso-sansone} DPT - diagrammatic
perturbation theory\cite{teichmann}) and analytical works (MFT -
mean-field theory,\cite{oktel} PA - Pad\'{e} analysis,\cite{niemeyer}
QRA - our calculations using quantum rotor approach).}

\label{comparison}
\end{table}
The transition between the SF and MI phases is associated with the
loss of long-range order. Let us introduce the notation for the maximum
of the critical value for parameter $t/U$ (as a function of the normalized
chemical potential $\mu/U$) at the tip of the first ($n_{B}=1$)
MI lobe for different lattices and frustration parameters $f$ as
follows \begin{equation}
x{}_{f}^{\square,\triangle}\equiv{\rm max}\left\{ \left(\frac{t}{U}\right)_{{\rm crit}}\right\} _{f}^{\square,\triangle}.\end{equation}
In Table \ref{comparison}, we compare values of the $x_{0}^{\square}$
resulting from several numerical\cite{capogrosso-sansone,teichmann}
and analytical studies.\cite{oktel,niemeyer} We found them in good
agreement however, mean-field theory calculations of the BH model
underestimate $x_{0}^{\square}$ and in the quantum rotor approach
there is a slight upward trend of the boundary towards higher critical
values of parameter $x_{0}^{\square}$ than obtained from numerical
calculations. The ground state of the rotating Bose-Einstein condensates
on a triangular lattice appears to be most stable against the effect
of rotation (see Fig. \ref{phase diagram 2dts}). The stability comes
from the higher values of the repulsive energy for the triangular
lattice. However, if the rotation frustration parameter is equal $f=1/3$
and $1/2$ the ratio $x_{f}^{\bigtriangleup}/x_{f}^{\square}$ of
the energy needed to cause loss of the global coherent state changes
character and is higher for triangular lattice unlike the cases with
$f=0,$ $1/6$ and $1/4$ (see Fig. \ref{phase diagram 2dts 12} and
Fig. \ref{triangular square phase}). In the above we choose $\mu/U=1/2$
because the transition at integer density belongs to the universality
class of the $2+1$ dimensional $XY$ model by contrast to transition
if one cross SF-MI phase boundary by variations in the chemical potential.
Behavior of the maximum repulsive energy $x_{f}^{\square,\triangle}$
in the rotating system with $f\neq0$ taken for special value of the
$\mu/U=1/2$ is non-monotonical in both square and triangular lattice
(Fig. \ref{square phase} and Fig. \ref{triangular phase-1}). While
critical values for $x_{f}^{\square,\triangle}$ are different for
various topologies of the system, the transition seen in the time-of-flight
images occurs rather rapidly with increasing lattice depth.\cite{greiner}
Because the experimental parameter $V_{0}/E_{R}$ ($V_{0}$ is the
maximum value of the lattice depth), depends logarithmically on $U/t$,
the small changes of the dimensionless depth of the optical lattice
can cover a wide range of the phase diagram. In order to verify the
calculated phase boundaries experimentally one shall be able to obtain
the higher resolution than required to distinguish the $n_{B}=1$
from $n_{B}=2$ transition with $x_{0}^{\square,\triangle}(n_{B}=2)/x_{0}^{\square,\triangle}\left(n_{B}=1\right)\approx0.65$.
With increasing number of particles per lattice site ($n_{B}\rightarrow$$\infty$),
the system possesses an exact particle-hole symmetry thus, there is
no difference between left $\bar{\mu}/U=n-\epsilon$ and right $\bar{\mu}/U=n+\epsilon$
branch of the $n$-th lobe, where $\epsilon\leq0.5$.


\section{Conclusions}

The physics of strongly correlated bosonic systems is the competition
between two tendencies of the bosons to spread out as a wave and to
localize as a particle combined with a frustration caused by rotation.
We presented a field-theoretic study of the ground-phase diagram in
quantum two-dimensional gaseous Bose-Einstein condensates where mentioned
emulation takes place. We calculated the phase diagram using the quantum
rotor approach with exactly evaluated density of states for two-dimensional
lattices with rational magnetic flux/rotation frustration parameter
$f=p/q$ for a number of values $f=1/q$. In systems that are in the
global coherent state at $f=0$, but with the ratio $t/U$ close to
the critical value $\left(t/U\right)_{\mathrm{crit}}$, a rotation
can be used to drive the condensates into the MI state (Fig. \ref{phase diagram 2dts}).
We compare the maximum of the critical value for $t/U$ parameter
(as a function of the normalized chemical potential $\mu/U$) at the
tip of the first ($n_{B}=1$) MI lobe for square lattice with several
numerical and analytical works and found them in a good agreement.
Note that the dependence of the $x_{f}^{\square,\triangle}/x_{0}^{\square,\triangle}$
from frustration parameter $f$ is non-monotonical (Fig. \ref{square phase}
and Fig. \ref{triangular phase-1}). The critical values of the energy
needed to drive a rotating condensate out of a global coherent state
change by varying the frustration parameter and strongly depends on
topology of the lattice.

The nice feature of presented approach, described in details above,
is that all the expressions and handling are analytic. It is also
worth to notice that we provide an exact formulas for density of states
that can be very useful in various situations whenever the magnetic
field/rotation is applied to the physical system. To our knowledge
the analytical expressions for DOS for triangular lattice were not
known in the literature. Notice, we consider only the limit $T\rightarrow0$
since in two-dimensional systems with a continuous symmetry the long-range
order is destroyed by the quantum fluctuations at finite temperature.\cite{mermin}
Moreover, we want to emphasize that our approach cannot be used for
analysis of the Berezinski-Kosterlitz-Thouless transitions since it
is appropriate only for physical systems where long-range order appears.
\begin{acknowledgments}
We thank R. Micnas and T. A. Zaleski for fruitful and stimulating
discussions and R. W. Chhajlany for careful reading of the manuscript.
\end{acknowledgments}
\appendix

\section{Density of States}

In this appendix we give the explicit formulas for the density of
states Eq. (\ref{DOS definition}) for square and triangular lattice
structures with uniform magnetic field/rotation. The provided analytical
expression can be advantageous in evaluating sums over momenta in
Eq. (\ref{critical line}). Moreover, the connection between the DOS
and Hofstadter butterfly will be shown.

\begin{figure}
\includegraphics[scale=0.48]{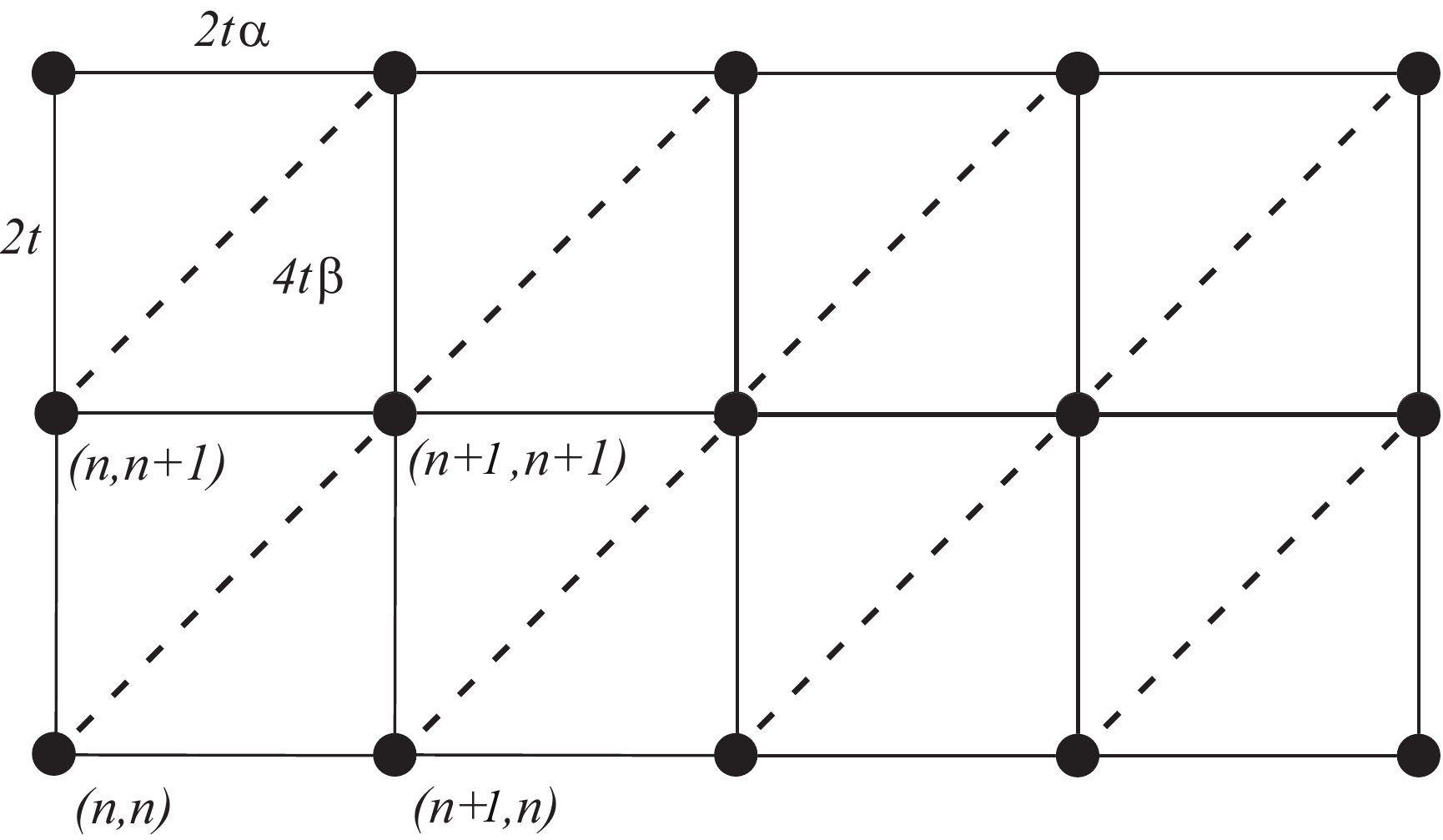}\caption{The lattice we used in our calculations is topologically equivalent
to triangular structure. It appears from square lattice when we add
bonds between next-nearest neighbors $\left(n,n\right)$ and $\left(n+1,n+1\right)$,
thus if $\phi$ is the flux per plaquette then for triangular lattice
we have $\phi/2$.}
\label{triangular topological}
\end{figure}
We start from the dispersion relevant for a lattice that can be viewed
as more general than square and triangular since it includes both
cases (Fig. \ref{triangular topological})\begin{equation}
t_{\boldsymbol{k}}=-2t\left[\cos k_{1}+\alpha\cos k_{2}+2\beta\cos\left(k_{1}+k_{2}\right)\right].\end{equation}
Regarding the above parameters $\alpha=0$ and $\beta=0$ lead to
$1D$ chain; $\alpha=1$ and $\beta=0$ to the square lattice\begin{equation}
t_{\boldsymbol{k}}^{\square}=-2t\left(\cos k_{1}+\cos k_{2}\right)\end{equation}
 and finally $\alpha=1$ and $\beta=0.5$ to the triangular structure\begin{equation}
t_{\boldsymbol{k}}^{\triangle}=-2t\left[\cos k_{1}+\cos k_{2}+\cos\left(k_{1}+k_{2}\right)\right].\end{equation}
 The different sign of the $\beta$ parameter changes the parity of
the DOS that will be very useful in some cases of triangular lattice
in rotating frame. Performing integration over momenta \begin{widetext}
\begin{equation}
\rho\left(\alpha,\beta,\xi\right)=\int_{-\pi}^{+\pi}\frac{d^{2}\mathbf{k}}{\left(2\pi\right)^{2}}\delta\left[\xi-\cos k_{1}-\alpha\cos k_{2}-2\beta\cos\left(k_{1}+k_{2}\right)\right]\end{equation}
we get: \begin{equation}
\rho\left(\alpha,\beta,\xi\right)=\left\{ \begin{array}{ccc}
\frac{1}{\pi^{2}\sqrt{\alpha+\beta\xi}}\mathbf{K}\left(\sqrt{\frac{\left(1+\alpha\right)^{2}-\left(\xi-\beta\right)^{2}}{4\left(\alpha+\beta\xi\right)}}\right) & for & \begin{array}{c}
1+\alpha+\beta\geq\xi\geq1-\alpha-\beta\\
-1+\alpha-\beta\geq\xi>-1-\alpha+\beta\end{array}\\
\frac{2}{\pi^{2}\sqrt{\left(1+\alpha\right)^{2}-\left(\xi-\beta\right)^{2}}}\mathbf{K}\left(\sqrt{\frac{4\left(\alpha+\beta\xi\right)}{\left(1+\alpha\right)^{2}-\left(\xi-\beta\right)^{2}}}\right) & for & 1-\alpha-\beta\geq\xi\geq-1+\alpha-\beta\end{array}\right.,\end{equation}
\end{widetext} where $\mathbf{K}\left(k\right)=\int_{0}^{\pi}dx\left(1-k^{2}\sin^{2}x\right)^{-1/2}$
is the complete elliptic function of the first kind.\cite{abramovitz}

\section{Square lattice in rotating frame}

The effects of homogeneous magnetic field/rotation on particles have
many interesting features. While energy levels are quantized into
Landau levels in $2D$ uniform space, a very rich structure appears
e.g. Hofstadter butterfly\cite{hofstadter}, when the lattice geometry
is taken into account.\cite{hasegawa} If one uses the Landau gauge
$A\left(0,x,0\right)$ then the dispersion $t_{\boldsymbol{k}}^{\square}$
for a square lattice of spacing $a=1$ with rotation frustration parameter
$f=p/q$ is given by\begin{equation}
\mathrm{det}\left(\begin{array}{ccccc}
M_{1} & -e^{-ik_{1}} & \cdots & 0 & -e^{-ik_{1}}\\
-e^{ik_{1}} & M_{2} & -e^{-ik_{1}} & 0 & 0\\
\vdots & -e^{ik_{1}} & M_{3} & \ddots & 0\\
0 & 0 & \ddots & \ddots & -e^{-ik_{1}}\\
-e^{-ik_{1}} & 0 & 0 & -e^{ik_{1}} & M_{n}\end{array}\right)=0,\label{harper equation square}\end{equation}
where \begin{equation}
M_{n}=-t_{\boldsymbol{k}}^{\square}-2\cos\left(k_{2}+2\pi fn\right).\end{equation}
Equation (\ref{harper equation square}) is known as Harper\textquoteright{}s
equation and has been studied extensively. If integers $p$ and $q$
are chosen to represent the angular velocity (with no common factor
in $p$ and q), then the dependence on the wave vector always appears
through the generalized structure factor $\gamma_{n}^{\square}=\cos nk_{1}+\cos nk_{2}$.
The density of states given by Eq. (\ref{DOS definition}) can be
obtained by computing energy bands $t_{\boldsymbol{k}}^{\square}$
from the eigenvalue equation (\ref{harper equation square}) see Table
\ref{eigenvalue equation square}.

\begin{table}
\begin{longtable}{c|c|>{\centering}p{3in}}
\hline
\hline 
$p$ & $q$ & $t_{\boldsymbol{k}}\equiv t_{\boldsymbol{k}}^{\square}$\tabularnewline
\endhead
\hline 
$1$ & 6 & $t_{\boldsymbol{k}}^{6}-12t_{\boldsymbol{k}}^{4}+24t_{\boldsymbol{k}}^{2}-4-2\gamma_{6}^{\square}=0$\tabularnewline
\hline
\hline
\endlastfoot
\hline 
$1$ & 2 & $t_{\boldsymbol{k}}^{2}-4-2\gamma_{2}^{\square}=0$\tabularnewline
\hline 
$1$ & 3 & $t_{\boldsymbol{k}}^{3}-6t_{\boldsymbol{k}}+2\gamma_{3}^{\square}=0$\tabularnewline
\hline 
$1$ & 4 & $t_{\boldsymbol{k}}^{4}-8t_{\boldsymbol{k}}^{2}+4-2\gamma_{4}^{\square}=0$\tabularnewline
\end{longtable}\caption{Energy dispersion for rotating square lattice for values $f=p/q$
used in calculations.}
\label{eigenvalue equation square}
\end{table}
The calculation of the exact formulae for DOS is straightforward,
although for large values of $q$ may only be done numerically. However,
for a number of $q$ values of interest it can be calculated analytically
with a closed-form expression for $\rho_{p/q}^{\square}\left(\xi\right)$
as the end result. Below we list these cases.

\subsection{Square lattice without rotation - $f=0$}

In the case of zero rotation the density of states for the square
$2D$ lattice reads simply\begin{equation}
\rho_{0}^{\square}\left(\xi\right)=\rho\left(1,0,\xi\right).\end{equation}

\subsection{Rotating square lattice with $f=1/2$}

\selectlanguage{polish}%
\inputencoding{latin2}\begin{equation}
\rho_{1/2}^{\square}\left(\xi\right)=\frac{\left|\xi\right|}{2}\rho_{0}^{\square}\left[\frac{1}{2}\left(\xi^{2}-4\right)\right]\Theta\left(8-\xi^{2}\right).\end{equation}
\inputencoding{latin9}\foreignlanguage{english}{In the above $\Theta\left(x\right)$
is the unit step function}\inputencoding{latin2}.\begin{widetext}

\selectlanguage{english}%

\subsection{Rotating square lattice with $f=1/3$}

\selectlanguage{polish}%
\begin{eqnarray}
\rho_{1/3}^{\square}\left(\xi\right) & = & \frac{1}{4\sqrt{2}}\left|\left(\xi^{2}-2\right)\sqrt{\xi^{2}-8}\right|\rho_{0}^{\square}\left[\frac{1}{2}\xi\left(\xi^{2}-6\right)\right]\nonumber \\
 & \times & \left\{ \left|\sec\left(\varphi+\frac{\pi}{2}\right)\right|\left[\Theta\left(\xi+1+\sqrt{3}\right)-\Theta\left(6-\xi^{2}\right)-\Theta\left(\xi-1-\sqrt{3}\right)\right]\right.\nonumber \\
 & + & \sec\left(\varphi+\frac{\pi}{6}\right)\left[\Theta\left(\xi+\sqrt{6}\right)-\Theta\left(\xi+2\right)+\Theta\left(\xi\right)-\Theta\left(\xi+1-\sqrt{3}\right)\right]\nonumber \\
 & + & \left.\sec\left(\varphi-\frac{\pi}{6}\right)\left[\Theta\left(\xi-1+\sqrt{3}\right)-\Theta\left(\xi\right)+\Theta\left(\xi-2\right)-\Theta\left(\xi-\sqrt{6}\right)\right]\right\} ,\end{eqnarray}

\begin{equation}
\varphi=\frac{1}{3}\arctan\left(\frac{\sqrt{32-\xi^{2}\left(\xi^{2}-6\right)^{2}}}{\xi\left(\xi^{2}-6\right)}\right).\end{equation}

\selectlanguage{english}%

\subsection{Rotating square lattice with $f=1/4$}

\selectlanguage{polish}%
\begin{eqnarray}
\rho_{1/4}^{\square}\left(\xi\right) & = & \frac{1}{2}\left|\xi^{2}-4\right|\rho_{0}^{\square}\left[\frac{1}{2}\left(\xi^{4}-8\xi^{2}+4\right)\right]\left\{ \sqrt{4+\left|\xi-4\right|}\left[\Theta\left(8-\xi^{2}\right)-\Theta\left(4+2\sqrt{2}-\xi^{2}\right)\right]\right.\nonumber \\
 & + & \left.\sqrt{4-\left|\xi^{2}-4\right|}\Theta\left(4-2\sqrt{2}-\xi^{2}\right)\right\} .\end{eqnarray}

\selectlanguage{english}%

\subsection{Rotating square lattice with $f=1/6$}

\selectlanguage{polish}%
\begin{eqnarray}
\rho_{1/6}^{\square}\left(\xi\right) & = & \frac{1}{4\sqrt[4]{2}}\left|\left(\xi^{4}-8\xi^{2}+8\right)\sqrt{16+8\xi^{2}-\xi^{4}}\right|\rho_{0}^{\square}\left[\frac{1}{2}\left(\xi^{6}-12\xi^{4}+24\xi^{2}-4\right)\right]\nonumber \\
 & \times & \left\{ \sqrt{2}\sec\left(\varphi-\frac{\pi}{2}\right)\sqrt{\cos\left[\frac{1}{2}\left(\varphi+\frac{\pi}{4}\right)\right]\cos\left[\frac{1}{2}\left(\varphi-\frac{\pi}{4}\right)\right]}\left[\Theta\left(5+\sqrt{21}-\xi^{2}\right)-\Theta\left(6+2\sqrt{3}-\xi^{2}\right)\right]\right.\nonumber \\
 & + & \sec\left(\varphi+\frac{\pi}{6}\right)\sqrt[4]{1-\sqrt{2}\cos\left(\varphi-\frac{\pi}{3}\right)-\frac{1}{2}\cos\left(2\varphi+\frac{\pi}{3}\right)}\Theta\left(5-\sqrt{21}-\xi^{2}\right)\nonumber \\
 & + & \left.\sec\left(\varphi-\frac{\pi}{6}\right)\sqrt[4]{1-\sqrt{2}\cos\left(\varphi+\frac{\pi}{3}\right)-\frac{1}{2}\cos\left(2\varphi-\frac{\pi}{3}\right)}\left[\Theta(6-2\sqrt{3}-\xi^{2})-\Theta(2-\xi^{2})\right]\right\} ,\end{eqnarray}
\begin{equation}
\varphi=\frac{1}{3}\arctan\left(\frac{\left|\left(\xi^{4}-8\xi^{2}+8\right)\sqrt{16+8\xi^{2}-\xi^{4}}\right|}{\xi^{6}-12\xi^{4}+24\xi^{2}+32}\right).\end{equation}

\selectlanguage{english}%
\inputencoding{latin9}%
\begin{figure}
\includegraphics[scale=0.7]{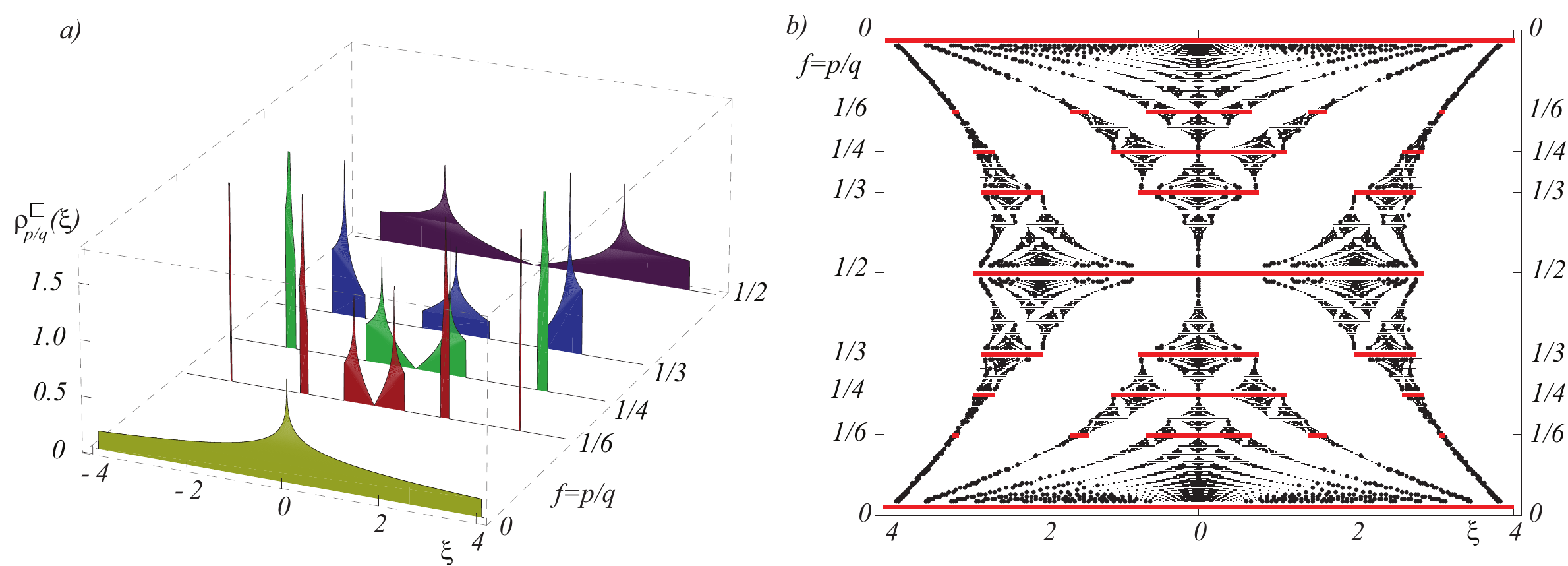}\caption{(Color online) a) The density of states for several values of the
frustration parameters $f=0$, $1/2$, $1/3$, $1/4$, and $1/6$
in square lattice. b) The projection of the DOS on $\xi-f$ surface
results in Hofstadter butterfly. Red lines correspond to our analytical
results.}
\label{dos 2dm}
\end{figure}

\selectlanguage{polish}%
\inputencoding{latin2}\end{widetext}\inputencoding{latin9}\foreignlanguage{english}{The
projection of the analytically calculated density of states for several
values of rotation on the $\xi-f$ surface results in the Hofstadter
butterfly (see Fig. \ref{dos 2dm}). The calculations of the Hofstadter
spectrum feasible for general purposes are impractical in applications
since the phase boundary calculated from Eq. (\ref{critical line final})
strongly depends on the structure of DOS.}

\selectlanguage{english}%

\section{Triangular lattice in rotating frame}

For the lattice topologically equivalent to triangular (Fig. \ref{triangular topological})
the Harper's equation takes form:

\begin{equation}
\mathrm{det}\left(\begin{array}{ccccc}
M_{1} & N_{1} & \cdots & 0 & \bar{N}_{n}\\
\bar{N}_{1} & M_{2} & N_{2} & 0 & 0\\
\vdots & \bar{N}_{2} & M_{3} & \ddots & 0\\
0 & 0 & \ddots & \ddots & N_{n-1}\\
N_{n} & 0 & 0 & \bar{N}_{n-1} & M_{n}\end{array}\right)=0,\end{equation}
where $n=q/2$ for even and $n=q$ for odd $q$. Moreover \begin{eqnarray}
M_{n} & = & -t_{\boldsymbol{k}}^{\triangle}-2\cos\left(k_{2}+4\pi fn\right),\\
N_{n} & = & -e^{-ik_{1}}-e^{2\pi f\left(2n+1\right)}e^{i\left(k_{1}+k_{2}\right)},\end{eqnarray}
and $\bar{N}_{n}$ is a complex conjugation. Now the generalized structure
factor can be written in form\begin{equation}
\gamma_{\pm n}^{\triangle}=-\cos nk_{1}-\cos nk_{2}\mp\cos\left(nk_{1}+nk_{2}\right).\end{equation}
From the above we can derive the equations for the energy dispersion
(see Table \ref{table triangular}) for several values of the frustration
rotation parameter. Analytical results sometimes come at a price of
the complexity of solutions and that is the case here. Therefore we
omit exact results for the dispersions $t_{\boldsymbol{k}}^{\triangle}$
and present them only in the simple case of $f=1/4$.

\begin{table}
\begin{longtable}{c|c|>{\centering}p{3in}}
\hline
\hline 
$p$ & $q$ & $t_{\boldsymbol{k}}\equiv t_{\boldsymbol{k}}^{\triangle}$\tabularnewline
\endhead
\hline 
$1$ & 6 & $t_{\boldsymbol{k}}^{3}-9t_{\boldsymbol{k}}+6-2\gamma_{-6}^{\triangle}=0$\tabularnewline
\hline
\hline
\endlastfoot
\hline 
$1$ & 1 & $t_{\boldsymbol{k}}+2\gamma_{1}^{\triangle}=0$\tabularnewline
\hline 
$1$ & 2 & $t_{\boldsymbol{k}}-2\gamma_{-2}^{\triangle}=0$\tabularnewline
\hline 
$1$ & 3 & $t_{\boldsymbol{k}}^{3}-9t_{\boldsymbol{k}}-6+2\gamma_{3}^{\triangle}=0$\tabularnewline
\hline 
$1$ & 4 & $t_{\boldsymbol{k}}^{2}-6-2\gamma_{4}^{\triangle}=0$\tabularnewline
\end{longtable}\caption{Energy dispersion for rotating triangular lattice for values $f=p/q$
used in calculations.}
\label{table triangular}
\end{table}

\subsection{Triangular lattice without rotation - $f=0$}

Cases with $f=0\left(1/3\right)$ and $f=1/2\left(1/6\right)$ are
relatively simple since the density of states can be easily obtained
by changing $\beta\rightarrow-\beta$ or equivalently $\xi\rightarrow-\xi$. 

\begin{equation}
\rho_{0}^{\triangle}\left(\xi\right)\equiv\rho\left(1,0.5,\xi\right).\end{equation}

\begin{figure}
\includegraphics[scale=0.77]{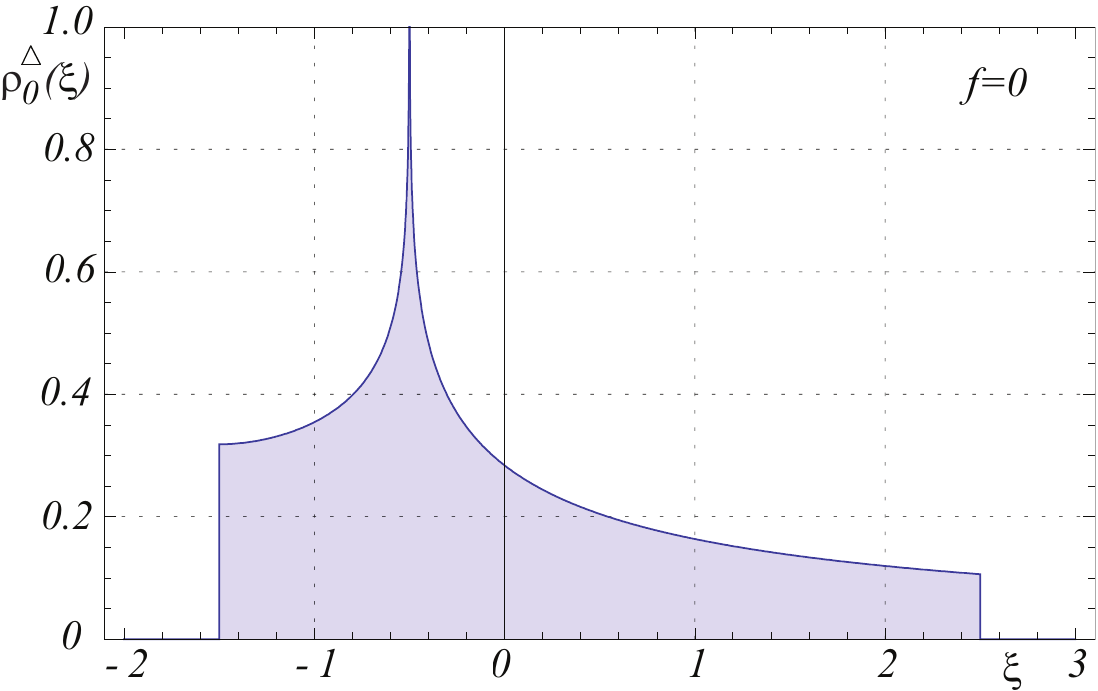}\caption{(Color online) DOS for triangular lattice without rotation $f=0$.}

\end{figure}

\subsection{Rotating triangular lattice with $f=1/2$}

\begin{equation}
\rho_{1/2}^{\triangle}\left(\xi\right)\equiv\rho\left(1,-0.5,\xi\right)=\rho_{0}^{\triangle}\left(-\xi\right).\end{equation}

\begin{figure}
\includegraphics[scale=0.77]{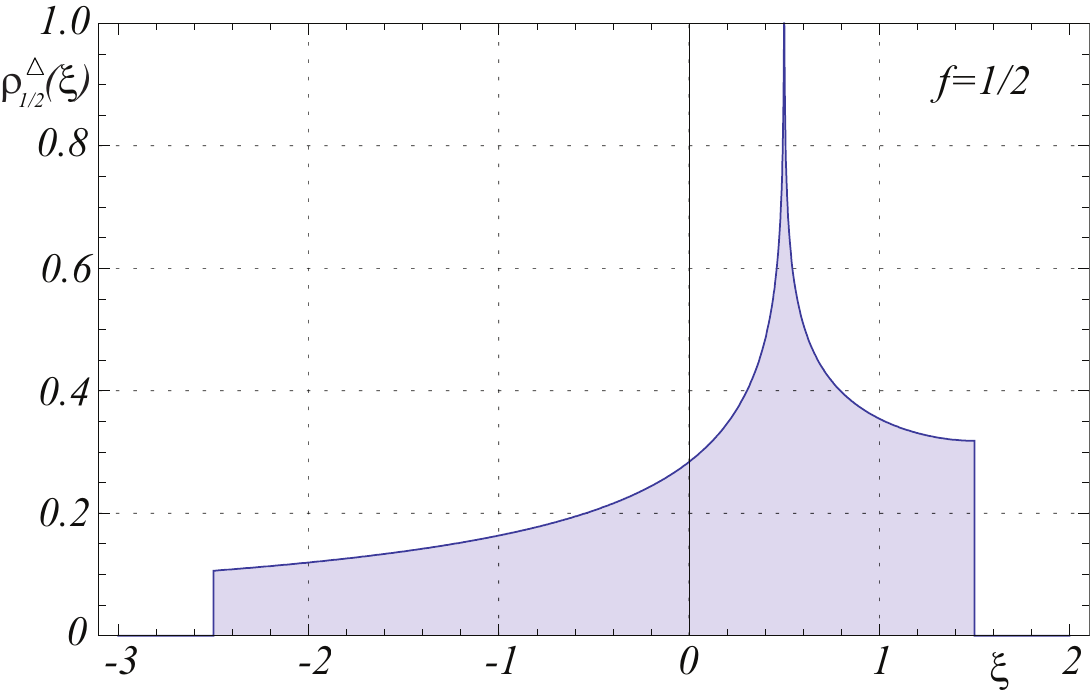}

\caption{(Color online) DOS for rotating triangular lattice $f=1/2$.}

\end{figure}

\subsection{Rotating triangular lattice with $f=1/3$}

\begin{widetext}

\begin{eqnarray}
\rho_{1/3}^{\triangle}\left(\xi\right) & = & 2^{1/3}\sqrt{\xi^{2}-12}\left|\left(\xi^{2}-3\right)\Phi^{2}\left(\xi\right)\right|\rho_{0}^{\triangle}\left[\frac{1}{2}\left(6+9\xi-\xi^{3}\right)\right]\nonumber \\
 & \times & \left\{ \left|\Phi\left(\xi\right)-6\right|^{-1}\left\{ \Theta\left[\xi-2\sqrt{3}\cos\left(\frac{\arctan\sqrt{107}}{3}\right)\right]-\Theta\left[\xi-2\sqrt{3}\cos\left(\frac{\pi}{18}\right)\right]\right\} \right.\nonumber \\
 & + & \left|-3\left(1-i\sqrt{3}\right)-\Phi\left(\xi\right)\right|^{-1}\left\{ \Theta\left[\xi+\theta\left(\frac{\arctan\sqrt{107}}{3}\right)\right]+\Theta\left[\xi+\theta\left(\frac{\pi}{18}\right)\right]\right\} \nonumber \\
 & + & \left.\left|6-\Phi\left(\xi\right)\right|^{-1}\left\{ \Theta\left[\xi+\theta\left(\frac{\arctan\sqrt{107}}{3}\right)\right]-\Theta\left[\xi+\theta\left(\frac{\pi}{18}\right)\right]\right\} \right\} ,\end{eqnarray}
\end{widetext}where\begin{equation}
\Phi\left(\xi\right)=2^{1/3}\left[\xi\left(\xi^{2}-9\right)+\left(\xi^{2}-3\right)\sqrt{\xi^{2}-12}\right]^{2/3},\end{equation}
and\begin{equation}
\theta\left(\varphi\right)=2\sqrt{3}\sin\varphi+\sqrt{2\sin^{2}\varphi-2\sqrt{3}\cos\varphi\sin\varphi+1}.\end{equation}
\begin{figure}
\includegraphics[scale=0.77]{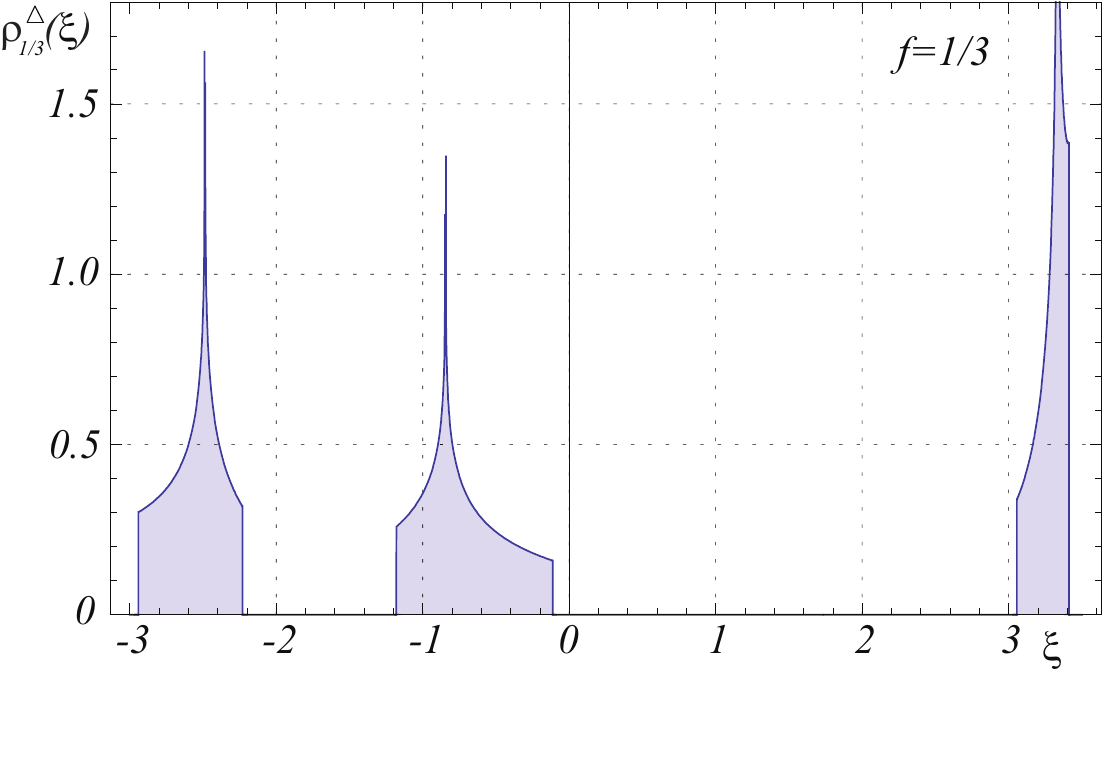}

\caption{(Color online) DOS for rotating triangular lattice $f=1/3$.}

\end{figure}

\subsection{Rotating triangular lattice with $f=1/4$}

The energy dispersion calculated from Harper's equation (\ref{harper equation square})
can be written in the form:\begin{equation}
t_{1/4}^{\triangle}=\left\{ \begin{array}{ccc}
+\sqrt{2\left(3+\gamma_{2}^{\triangle}\right)} & for & t_{\boldsymbol{k}}^{\triangle}\in\left[-\sqrt{11},0\right),\\
-\sqrt{2\left(3+\gamma_{2}^{\triangle}\right)} & for & t_{\boldsymbol{k}}^{\triangle}\in\left[0,\sqrt{11}\right].\end{array}\right.\end{equation}
Integrating over the wave vectors belonging to the Brillouin zone,
we obtain\begin{equation}
\rho_{1/4}^{\triangle}\left(\xi\right)=\frac{\left|\xi\right|}{2}\rho_{0}^{\triangle}\left[-\frac{1}{2}\left(\xi^{2}-6\right)\right].\end{equation}
\begin{figure}
\includegraphics[scale=0.77]{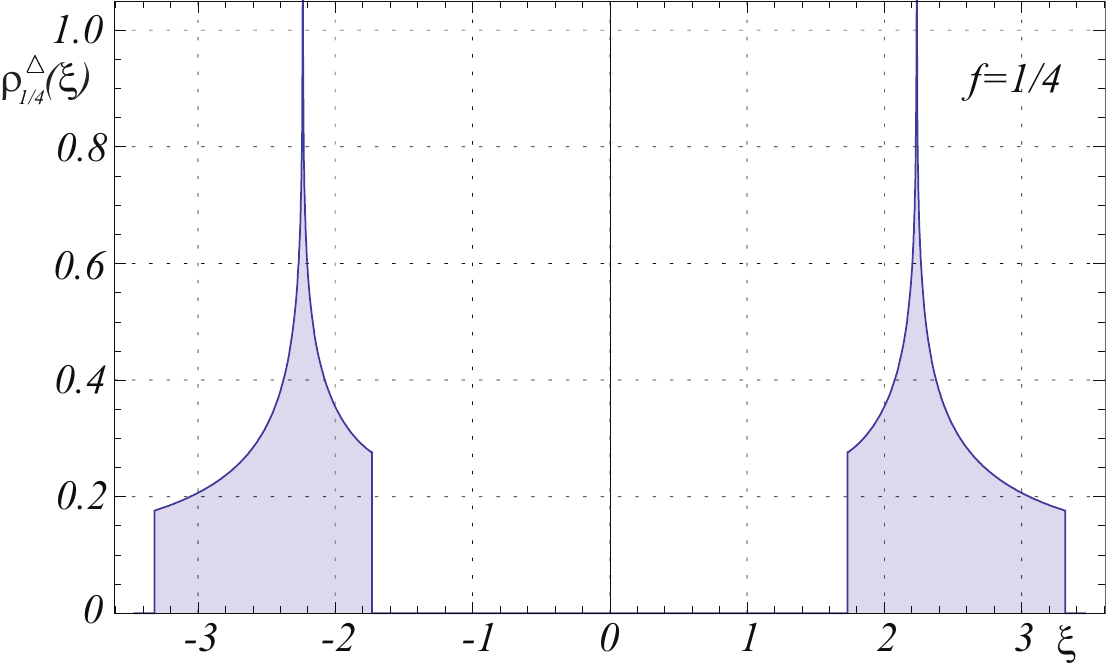}

\caption{(Color online) DOS for rotating triangular lattice $f=1/4$.}

\end{figure}

\subsection{Rotating triangular lattice with $f=1/6$}

\begin{equation}
\rho_{1/6}^{\triangle}\left(\xi\right)=\rho_{1/3}^{\triangle}\left(-\xi\right).\end{equation}

\begin{figure}
\includegraphics[scale=0.77]{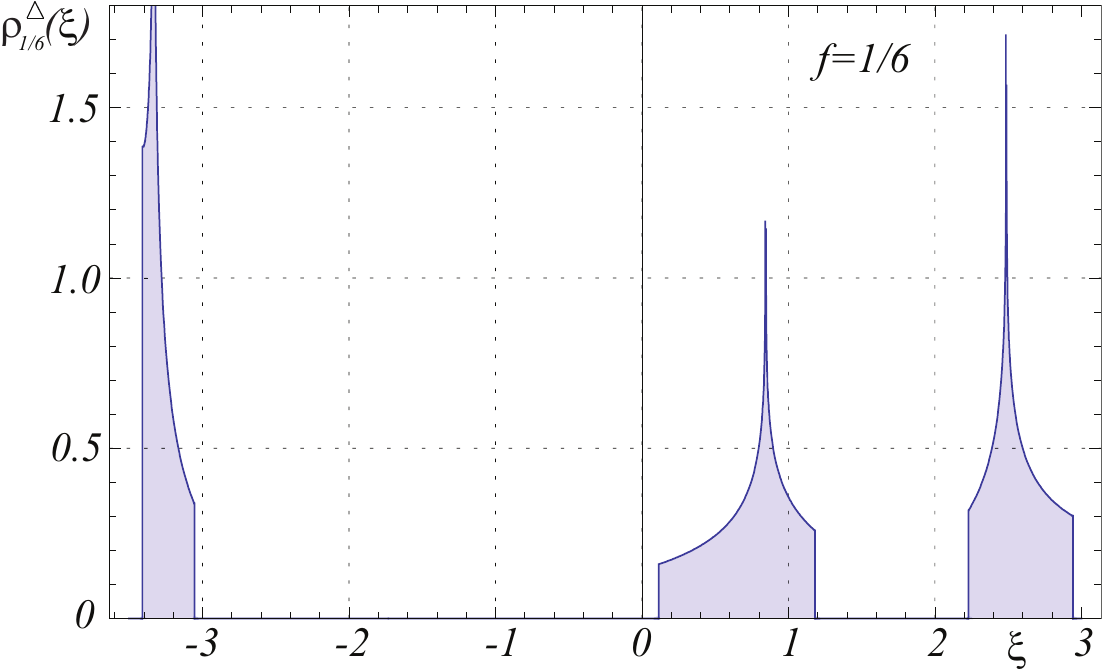}\caption{(Color online) DOS for rotating triangular lattice $f=1/6$.}

\end{figure}

\end{document}